\documentclass[a4paper]{jpconf}

\usepackage{graphicx}
\usepackage{color}
\usepackage{subfig}
\usepackage{amssymb}
\usepackage{amsmath}
\newcommand{\G}{\mathcal{G}}
\newcommand{\Gbar}{\bar{\mathcal{G}}}

\begin{document}

%% Title block
\title{Free-flight experiments in LISA Pathfinder}

% jpconf class is *incompatible* with authblk package. have to do it the old-fashioned way!
% Author list 
\author{
M~Armano$^{a}$,
H~Audley$^{b}$,
G~Auger$^{c}$,
J~Baird$^{n}$,
P~Binetruy$^{c}$,
M~Born$^{b}$,
D~Bortoluzzi$^{d}$,
N~Brandt$^{e}$,
A~Bursi$^{t}$,
M~Caleno$^{f}$,
A~Cavalleri$^{g}$,
A~Cesarini$^{g}$,
M~Cruise$^{h}$,
C~Cutler$^{u}$,
K~Danzmann$^{b}$, 
I~Diepholz$^{b}$, 
R~Dolesi$^{g}$,
N~Dunbar$^{i}$,
L~Ferraioli$^{j}$,
V~Ferroni$^{g}$,
E~Fitzsimons$^{e}$,
M~Freschi$^{a}$,
J~Gallegos$^{a}$,
C~Garc\'ia Marirrodriga$^{f}$,
R~Gerndt$^{e}$,
LI~Gesa$^{k}$,
F~Gibert$^{k}$,
D~Giardini$^{j}$,
R~Giusteri$^{g}$,
C~Grimani$^{l}$,
I~Harrison$^{m}$,
G~Heinzel$^{b}$, 
M~Hewitson$^{b}$, 
D~Hollington$^{n}$,
M~Hueller$^{g}$,
J~Huesler$^{f}$,
H~Inchausp\'e$^{c}$,
O~Jennrich$^{f}$,
P~Jetzer$^{o}$,
B~Johlander$^{f}$,
N~Karnesis$^{k}$,
B~Kaune$^{b}$,
N~Korsakova$^{b}$,
C~Killow$^{p}$,
I~Lloro$^{k}$,
R~Maarschalkerweerd$^{m}$,
S~Madden$^{f}$,
P~Maghami$^{s}$,
D~Mance$^{j}$,
V~Mart\'{i}n$^{k}$,
F~Martin-Porqueras$^{a}$,
I~Mateos$^{k}$,
P~McNamara$^{f}$,
J~Mendes$^{m}$,
L~Mendes$^{a}$,
A~Moroni$^{t}$,
M~Nofrarias$^{k}$,
S~Paczkowski$^{b}$,
M~Perreur-Lloyd$^{p}$,
A~Petiteau$^{c}$, 
P~Pivato$^{g}$,
E~Plagnol$^{c}$, 
P~Prat$^{c}$,
U~Ragnit$^{f}$,
J~Ramos-Castro$^{q}$$^{r}$,
J~Reiche$^{b}$,
J~A~Romera Perez$^{f}$,
D~Robertson$^{p}$, 
H~Rozemeijer$^{f}$,
G~Russano$^{g}$,
P~Sarra$^{t}$,
A~Schleicher$^{e}$,
J~Slutsky$^{s}$,
C~F~Sopuerta$^{k}$,
T~Sumner$^{n}$, 
D~Texier$^{a}$,
J~Thorpe$^{s}$,
C~Trenkel$^{i}$,
H~B~Tu$^{g}$,
D~Vetrugno$^{g}$,
S~Vitale$^{g}$,
G~Wanner$^{b}$, 
H~Ward$^{p}$,
S~Waschke$^{n}$,
P~Wass$^{n}$, 
D~Wealthy$^{i}$,
S~Wen$^{g}$,
W~Weber$^{g}$,
A~Wittchen$^{b}$,
C~Zanoni$^{d}$,
T~Ziegler$^{e}$,
P~Zweifel$^{j}$}

\address{$^{a}$ European Space Astronomy Centre, European Space Agency, Villanueva de la
Ca\~{n}ada, 28692 Madrid, Spain}
\address{$^{b}$ Albert-Einstein-Institut, Max-Planck-Institut f\"ur
Gravitationsphysik und Universit\"at Hannover, 30167 Hannover, Germany}
\address{$^{c}$ APC UMR7164, Universit\'e Paris Diderot, Paris, France}
\address{$^{d}$ Department of Industrial Engineering, University of Trento, via Sommarive 9, 38123 Trento, 
and Trento Institute for Fundamental Physics and Application / INFN}
\address{$^{e}$ Airbus Defence and Space, Claude-Dornier-Strasse, 88090 Immenstaad, Germany}
\address{$^{f}$ European Space Technology Centre, European Space Agency, 
Keplerlaan 1, 2200 AG Noordwijk, The Netherlands}
\address{$^{g}$ Dipartimento di Fisica, Universit\`a di Trento and Trento Institute for 
Fundamental Physics and Application / INFN, 38123 Povo, Trento, Italy}
\address{$^{h}$ Department of Physics and Astronomy, University of
Birmingham, Birmingham, UK}
\address{$^{i}$ Airbus Defence and Space, Gunnels Wood Road, Stevenage, Hertfordshire, SG1 2AS, UK }
\address{$^{j}$ Institut f\"ur Geophysik, ETH Z\"urich, Sonneggstrasse 5, CH-8092, Z\"urich, Switzerland}
\address{$^{k}$ Institut de Ci\`encies de l'Espai (CSIC-IEEC), Campus UAB, Facultat de Ci\`encies, 08193 Bellaterra, Spain}
\address{$^{l}$ Istituto di Fisica, Universit\`a degli Studi 
di Urbino/ INFN Urbino (PU), Italy}
\address{$^{m}$ European Space Operations Centre, European Space Agency, 64293 Darmstadt, Germany }
\address{$^{n}$ The Blackett Laboratory, Imperial College London, UK}
\address{$^{o}$ Physik Institut, 
Universit\"at Z\"urich, Winterthurerstrasse 190, CH-8057 Z\"urich, Switzerland}
\address{$^{p}$ SUPA, Institute for Gravitational Research, School of Physics and Astronomy, University of Glasgow, Glasgow, G12 8QQ, UK}
\address{$^{q}$ Department d'Enginyeria Electr\`onica, Universitat Polit\`ecnica de Catalunya,  08034 Barcelona, Spain}
\address{$^{r}$ Institut d'Estudis Espacials de Catalunya (IEEC), C/ Gran Capit\`a 2-4, 08034 Barcelona, Spain}
\address{$^{s}$ NASA Goddard Space Flight Center, 8800 Greenbelt Road, Greenbelt, MD 20771, USA}
\address{$^{t}$ CGS S.p.A, Compagnia Generale per lo Spazio, Via Gallarate, 150 - 20151 Milano, Italy}
\address{$^{t}$ NASA Jet Propulsion Laboratory, 4800 Oak Grove Drive, Pasadena, CA 91109}

\ead{james.i.thorpe@nasa.gov}

%% Abstract
\begin{abstract}
The LISA Pathfinder mission will demonstrate the technology of drag-free test masses for use as inertial references in future space-based gravitational wave detectors. To accomplish this, the Pathfinder spacecraft will perform drag-free flight about a test mass while measuring the acceleration of this primary test mass relative to a second reference test mass. Because the reference test mass is contained within the same spacecraft, it is necessary to apply forces on it to maintain its position and attitude relative to the spacecraft. These forces are a potential source of acceleration noise in the LISA Pathfinder system that are not present in the full LISA configuration. While LISA Pathfinder has been designed to meet it's primary mission requirements in the presence of this noise, recent estimates suggest that the on-orbit performance may be limited by this `suspension noise'. The drift-mode or free-flight experiments provide an opportunity to mitigate this noise source and further characterize the underlying disturbances that are of interest to the designers of LISA-like instruments. This article provides a high-level overview of these experiments and the methods under development to analyze the resulting data. 
\end{abstract}

%% INTRO
\section{Introduction}
\label{sec:intro}

The basic operating principle of an interferometric gravitational-wave detector is the measurement of fluctuations in space-time curvature via the exchange of photons between pairs of geodesic-tracking references separated by large baselines \cite{Bondi1959}.  A key challenge for implementing such a detector is the development of an object whose worldline approximates a geodesic -- an \emph{inertial} particle. The LISA Pathfinder (LPF) mission \cite{Armano_09} will validate the technology of drag-free test masses for use as inertial references in a future space-based gravitational wave detector such as the Laser Interferometer Space Antenna (LISA). 

The technique of drag-free flight for disturbance reduction \cite{Lange1964, DeBra1997} can be briefly summarized as follows. A reference or `test' mass is placed inside a hollow housing within the host spacecraft (SC). On orbit, the test mass is allowed to float freely inside the housing while a sensor system monitors the position and attitude of the test mass relative to the SC. This information is used by a control system which commands the SC to follow the orbit of the test mass. In doing so, the SC isolates the test mass from external disturbances. By actuating the SC rather than the test mass (as is done in a traditional inertial guidance system) to maintain the relative position and attitude, the test mass is isolated from noise associated with the actuation itself. The remaining residual acceleration noise of the test mass results from forces local to the SC as well as external forces that are not absorbed by the SC (e.g. magnetic). In practice, the ``accelerometer'' (feedback to the test mass) and ``drag-free'' (feedback to the SC) techniques are combined into a hybrid system where the feedback can be routed differently depending on the kinematic degree of freedom (DoF) and frequency band. For the LISA application, it is sufficient to have drag-free flight only along the linear DoF and only in the frequency band of desired sensitivity. This is analogous to the pendulum suspensions used for ground-based gravitational wave detectors that provide approximate free-fall in one DoF for frequencies sufficiently far above the resonance frequency.

In the LPF implementation \cite{Antonucci2011}, two $46\,\mbox{mm}$ cubic Au-Pt test masses are contained in a single SC. One of the test masses is designated as the reference test mass (RTM) and the SC performs drag free flight along one linear DoF while controlling the remaining five DoFs (three angular and two linear) with an electrostatic suspension system. The second or non-reference test mass (NTM) is located $\sim 38\,\mbox{cm}$ away from the primary test mass along the drag-free axis and is used as a witness to assess the residual acceleration of the RTM. An interferometric metrology system \cite{Audley2011}, monitors relative displacements between the RTM and NTM with a precision of $\sim10\,\mbox{pm}/\sqrt{\mbox{Hz}}$ in the measurement band $0.1\,\mbox{mHz}< f < 100\,\mbox{mHz}$

\begin{figure}[h!]
	\centering
	\subfloat[Design Requirements\label{fig:LPFreqs}]{\includegraphics[width=7 cm]{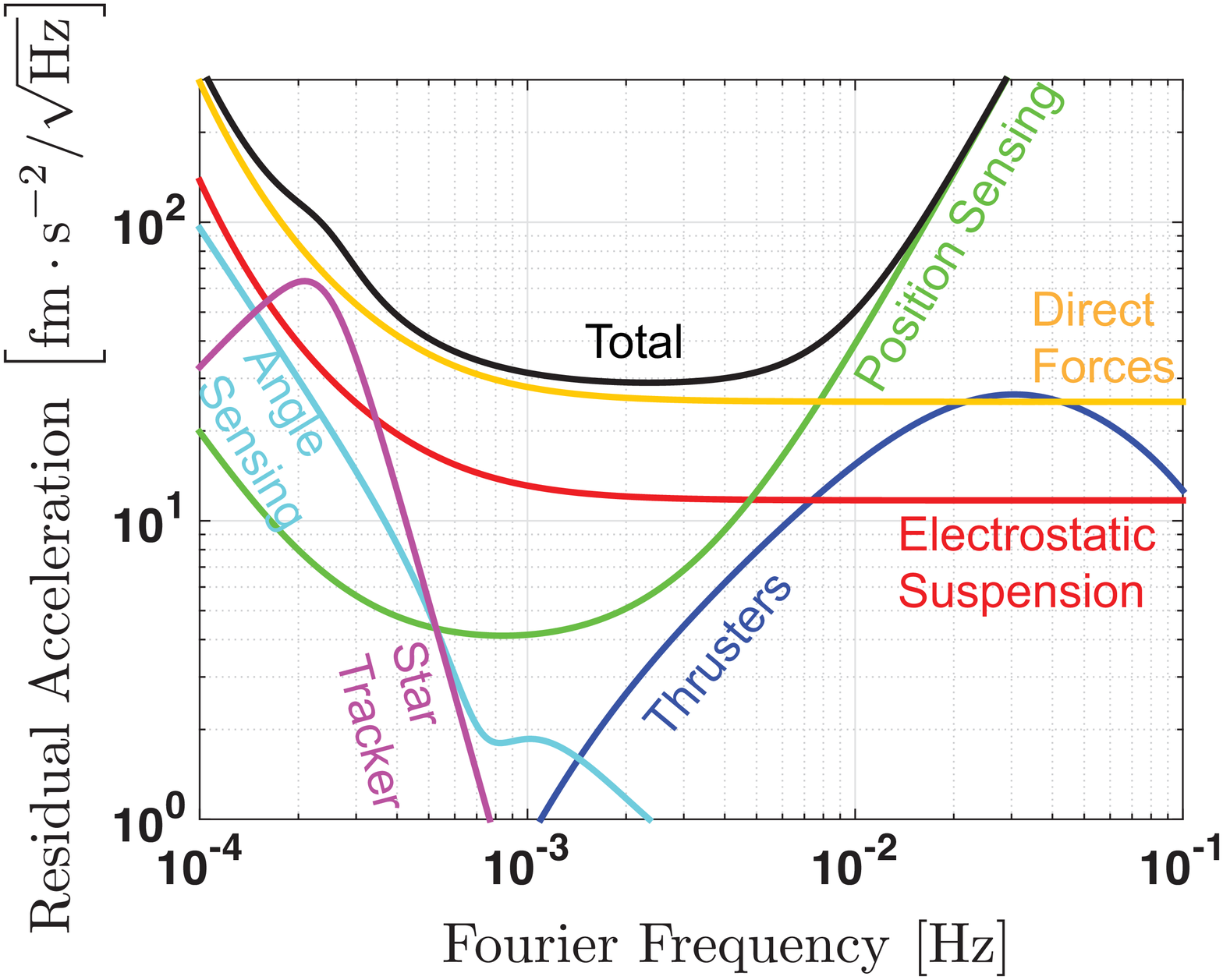}}\quad
	\subfloat[Current Best Estimate\label{fig:LPFCBE}]{\includegraphics[width=7 cm]{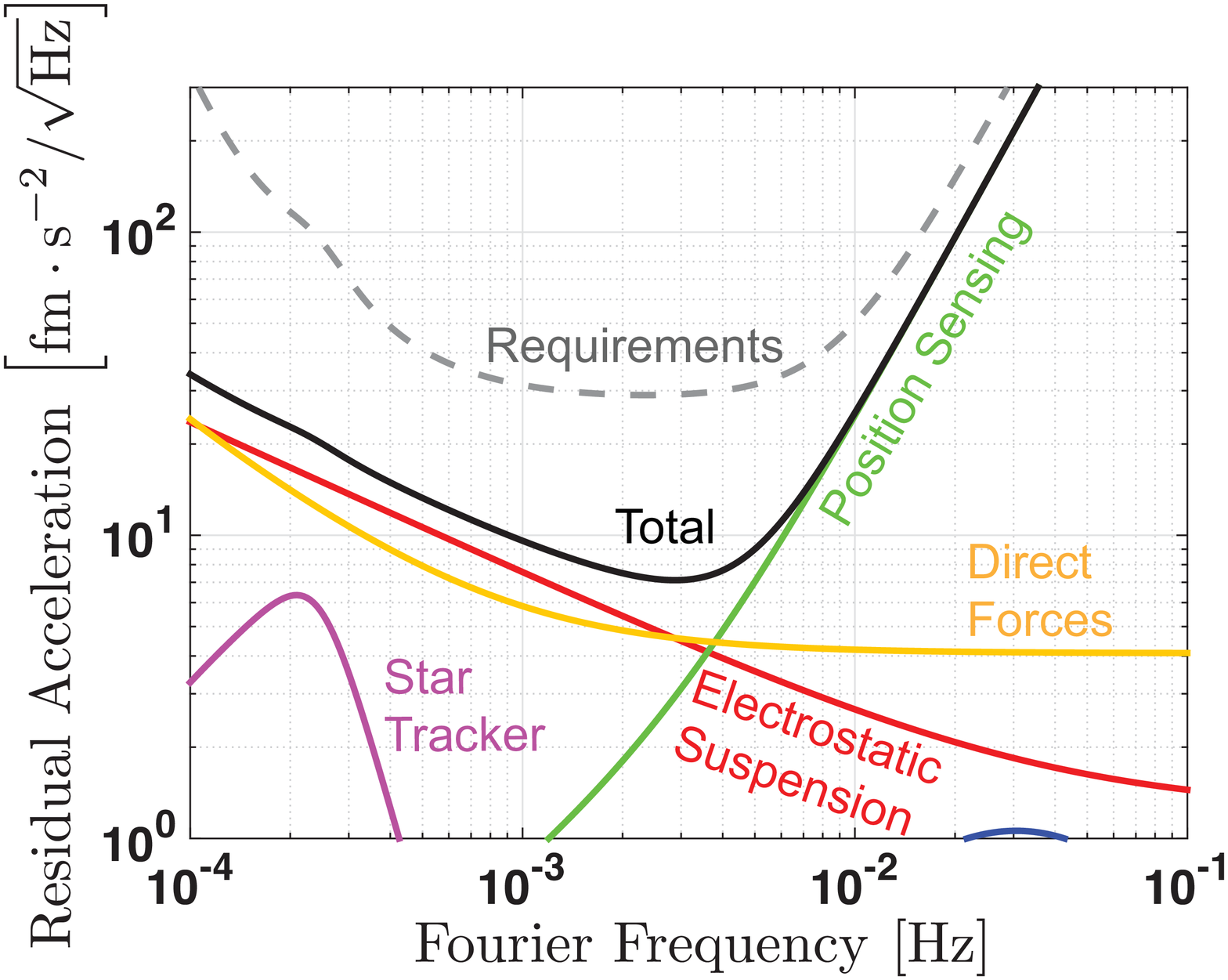}}\\	
	\caption{Breakdown of noise sources for measurement of acceleration of the reference test mass relative to the non-reference test mass along the sensitive DoF. The left panel shows the design requirement levels whereas the right panel shows the current best estimates based on ground test campaigns of flight hardware and system modeling. See \cite{Antonucci2011} for more detail.}
\end{figure}

A key difference between the LPF configuration and the LISA configuration is that the LISA acceleration measurement is made between test masses on \emph{separate} spacecraft whereas the LPF acceleration measurement is made between two test masses on the \emph{same} spacecraft. As a result, the LPF NTM must be electrostatically suspended along the sensitive DoF because the SC cannot simultaneously follow the trajectories of both the RTM and NTM along the same DoF.  This suspension force has a noise component which represents a disturbance in the LPF measurement that is not present in LISA-like configurations. With all noise sources at the design requirement levels for LPF, this residual suspension noise contribution is not a significant contribution to the overall measured differential acceleration noise in the LISA measurement band (see Figure \ref{fig:LPFreqs}).  The dominating term over this band are various types of spurious forces on the test masses, precisely the phenomena of interest to the designers of future gravitational wave instruments.

Figure \ref{fig:LPFCBE}  shows the current best estimate (CBE) for the differential acceleration noise in LPF.  As would be expected from a conservative set of design requirements, the CBE levels for all terms are lower than the corresponding ones for the requirements. However, suspension noise is now expected to play a significant role in the measurement band. This expectation is supported by stringent limits placed on the magnitudes of unknown or unmodeled forces on the LPF inertial sensor using torsion pendulums \cite{Carbone2007, Cavalleri2009}.

To be clear, LPF performance at either the requirement or CBE level would accomplish the goal of validating drag-free flight as a technique for realizing inertial reference sensors for LISA-like observatories. Nevertheless, a direct measurement of the forces acting on the test masses would provide additional valuable information to the designers of such observatories.

\section{Free-Flight Experiments}
\label{sec:DMconcept}

The electrostatic suspension of the NTM in LPF is needed to counteract forces in the NTM-SC system that differ from those in the RTM-SC system, which are  suppressed by the drag-free control loop. These include disturbance forces that are local to the test masses, such as residual gas disturbances, thermal disturbances, electrostatic forces, magnetic forces, etc. In addition, the static gravitational gradient differs at the RTM and NTM location, leading to a constant bias in the force that must be applied to the NTM along the sensitive axis. Although this bias will be minimized through the use of compensation masses designed based on pre-flight gravitational models, it is expected that residual gravitational accelerations along the sensitive DoF will be on the order of $10^{-10}\,\mbox{m}/\mbox{s}^2$. This sets the amplitude, and consequently the noise, of the required suspension force.

In the \emph{drift mode} or \emph{free-flight} experiment, the compensation of the static field experienced by the NTM is performed with a series of discrete `kicks' rather than with a continuously-applied force. Between the kicks, the electrostatic actuation of the NTM along the sensitive DoF is turned off, allowing the NTM to drift under the influence of the constant forces. In principle, this `kick control' strategy could be employed on all DoFs of the NTM, which would suppress actuation noise from other DoFs from leaking into the sensitive DoF. In practice, this actuation cross-talk is expected to be sufficiently small that kick control is only required along the sensitive DoF. 

\subsection{The LTP Drift Mode Experiment}
\begin{figure}[h!]
	\centering
	\subfloat[RTM-NTM Relative Position\label{fig:timeseries_X}]{\includegraphics[width=7 cm]{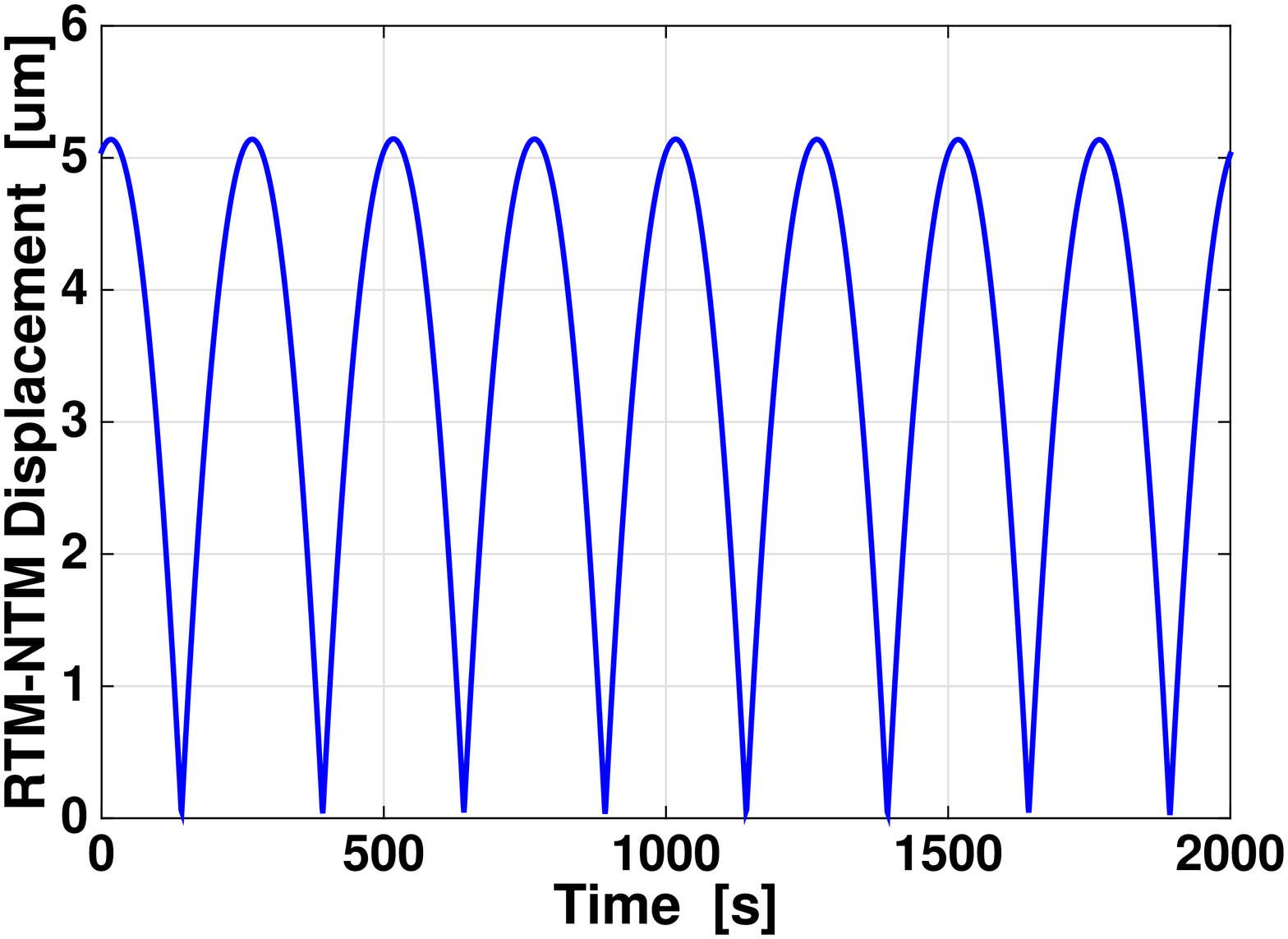}}\quad
	\subfloat[RTM-NTM Relative Acceleration\label{fig:timeseries_A}]{\includegraphics[width=7 cm]{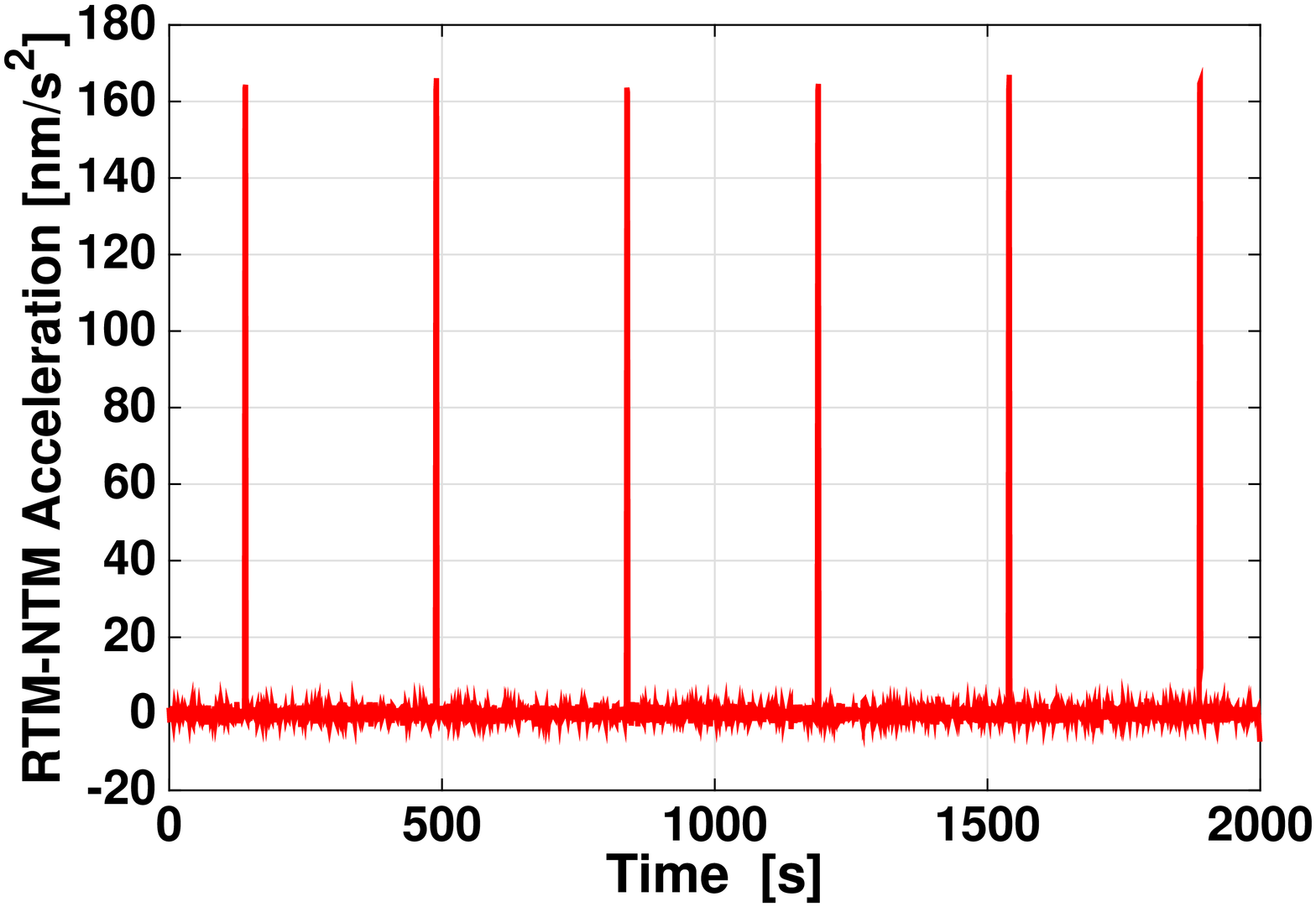}}\\	
	\caption{Example data for a LTP drift-mode simulation. The left panel shows the displacement of the RTM relative to the NTM whereas the right panel shows the acceleration.}
\end{figure}

The drift mode controller designed for the LISA Test Package (LTP)\cite{Grynagier10} fixes both the length of the drift period ($\sim 350\,\mbox{s}$) and the duration of the kicks ($\sim 1\,\mbox{s}$). A Kalman-filter based observer tracks the motion of the NTM during the free flight and estimates the impulse required to maintain the NTM position. The amplitude of the subsequent kick is adjusted to deliver this impulse and the process is repeated. Figure \ref{fig:timeseries_X} shows the displacement of the RTM relative to the NTM for a segment of simulated data from the LTP drift mode. It consists of a series of repeated quasi-parabolic flights (the trajectory of the NTM is not a true parabola due to the influence of other force terms such as a linear spring term that couples the NTM to the SC) with a duration of $350\,\mbox{s}$. Figure \ref{fig:timeseries_A} shows the corresponding acceleration of the RTM relative to the NTM, which shows a series of discrete kicks separated by nearly constant acceleration during the free flight segments.

\subsection{A proposed ST7-DRS Free-Flight Experiment}
An alternative drift-mode control design has recently been develop for possible use with the NASA-provided payload known as the Disturbance Reduction System (DRS)\footnote{the DRS is part of NASA's Space Technology 7 (ST7) mission}. The ST7 design is based on a modified dead-band controller, where the dynamical state of the system determines the control mode.  The two-dimensional phase space of NTM position and NTM velocity is used to describe the system state and a region centered around the nominal position and zero velocity is designated as the dead band. So long as the NTM position and velocity remain in this dead band region, no suspension control is applied along the sensitive DoF.  When the NTM state drifts out of this deadband region, a suspension controller is engaged and remains on until the NTM state re-enters the dead-band region.  The resulting trajectory of the NTM is more complex than that of the LTP drift mode experiment. In general, the duration of the free-flight segments vary from segment to segment. A 300 ks simulation yielded an approximately normal distribution of free-flights with a mean duration of $\sim 357\,\mbox{s}$ and a standard deviation of  $\sim 14\,\mbox{s}$. While at first glance, this irregularity in the free flight durations might seem to be disadvantage, it has a potential advantage for data analysis because the few long segments can be used to estimate spectral information at lower frequencies. Additionally, unlike the LTP design,  the DRS control design automatically adjusts to changes in the static gravity gradient and is robust to fairly significant changes. This means that the controller would not have to be re-optimized in-flight as the residual gravity gradient changes, for example due to fuel consumption.

\subsection{Laboratory Experiments}
\label{sec:lab}

Torsion pendulums with a LPF-like test mass suspended as a torsion member have been used to measure the small forces relevant to the free-fall purity in LISA and LPF.  In this section, we describe an implementation of free-flight experiments in a torsion pendulum facility located at the University of Trento. The torsion pendulum consists of a hollow replica of the LISA Pathfinder test mass, suspended by a thin silica torsion fiber, and hangs inside a Gravitational Reference Sensor (GRS) prototype \cite{Cavalleri2009}.  The sensitive degree of freedom is $\Phi$, the rotation around the z-axis. A breadboard version of the front-end electronic chain provides torque authority of $\sim 200\,\mbox{fN}\cdot\mbox{m}$ by applying desired actuation voltages acrossa diagonal pair of electrodes. To simulate a large DC acceleration, the pendulum can be rotated by an angle $\Delta \Phi$ with respect to the inertial sensor, such that a DC torque \begin{math} N_{DC} = - 
\end{math} $\Gamma$ \begin{math}\cdot\end{math} $\Delta \Phi$ is required to keep it centered. This is analogous to the bias on the NTM suspension in LPF. Unlike LPF, the bias for the torsion pendulum can be tuned by adjusting $\Delta\Phi$.

The equation of motion for the torsional degree of freedom for the test-mass is
 \begin{equation}
 I\ddot{\Phi}=-I\omega^{2}_{0}\Phi-\frac{I}{\tau}\dot{\Phi}+N(t)
 \end{equation}
where $I$ is the moment of inertia, \begin{math}\omega_{0}=\sqrt{\frac{\Gamma}{I}}=\frac{2\pi}{T_{0}}\end{math} is the pendulum resonance angular frequency (and $T_{0}$ the period), $\Gamma$ is the pendulum rotational elastic constant and $\tau$ is the energy decay time. 
It is possible to soften electrostatically the pendulum by applying DC constant voltages to lengthen the pendulum period from roughly $465\,\mbox{s}$, without applied fields, to as much as \begin{math}T_{0}\approx 830\,\mbox{s}\end{math}, to allow flight times comparable to those foreseen for LPF, \begin{math}T_{fly}=350\,\mbox{s}\end{math}.

The pendulum torque sensitivity is around $1\,\mbox{fN}\cdot\mbox{m}/\sqrt{\mbox{Hz}}$ at $1\,\mbox{mHz}$, corresponding to an equivalent acceleration of $50\,\mbox{fm}/\mbox{s}^2 \sqrt{\mbox{Hz}}$, and thus near the LPF noise specification.  The pendulum can thus allow a quantitatively significant test of the free-fall mode and our analysis algorithms, with real data exhibiting gaps as well as the large dynamic range needed in free-fall mode, and thus readout and dynamic system linearity challenges.  The on-ground experiment will also allow more flexibility to explore different control strategies, by varying flight and impulse time or control points, and different dynamic configurations made possible by having a variable stiffness. 

Our free-fall test consists of three measurements. The first is the pendulum background torque noise level in absence of any applied force, which can be performed by rotating the pendulum such that the test mass is centered without applied torques. The measured angular displacement is then converted into torque

 \begin{equation}
 N_{m}=I\ddot{\Phi}+\Gamma\Phi+\frac{I}{\tau}\dot{\Phi}
 \label{eq:rot2torque}
 \end{equation} 

\begin{figure}[h!]
	\centering
	\subfloat[Torsion-pendulum apparatus\label{fig:utn_pendulum_pic}]{\includegraphics[width=7 cm]{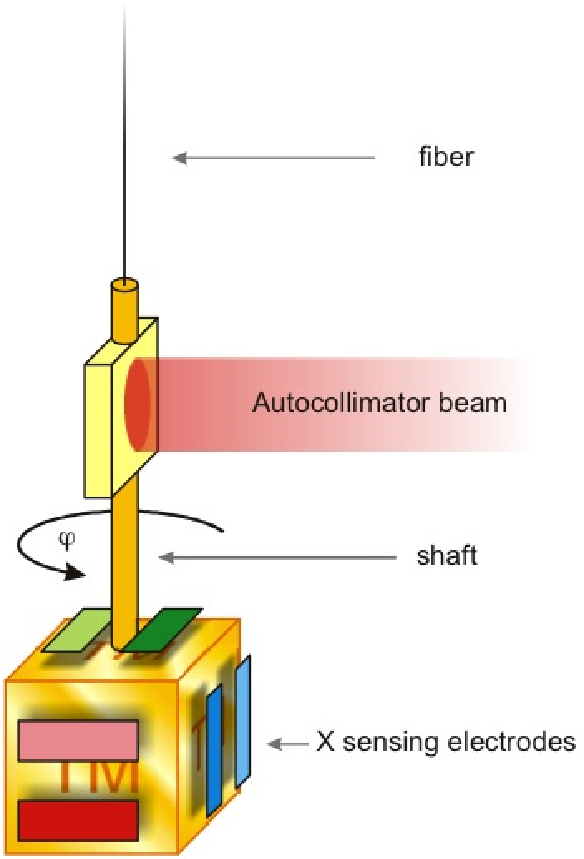}}\quad
	\subfloat[Free-flight displacement data\label{fig:utn_pendulum_data}]{\includegraphics[width=7 cm]{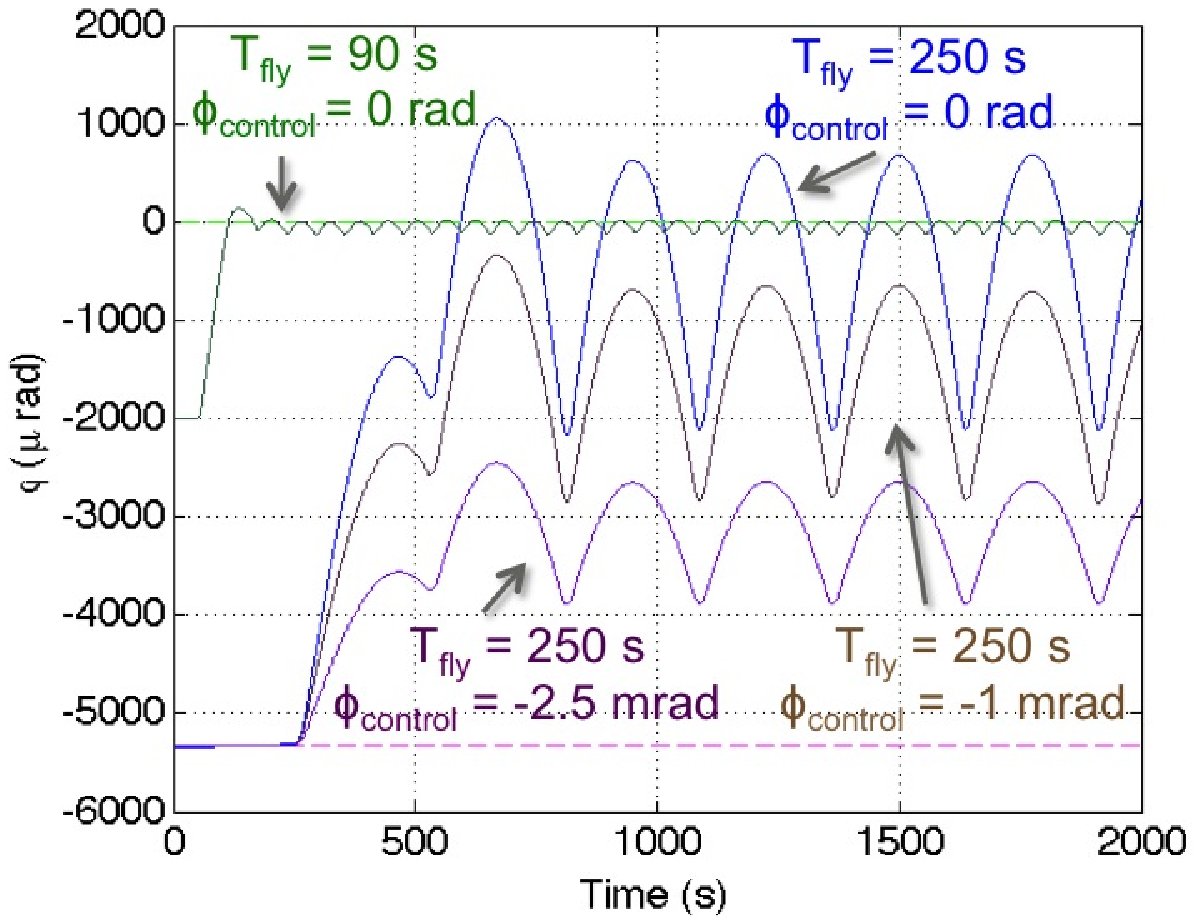}}\\	
	\caption{Example data from a torsion pendulum facility used to simulate LPF free-flight experiments. The displacement timeseries is for an experiment performed with free-flight times $T_{fly} = 90\,\mbox{s}$ and $250\,\mbox{s}$ and angular set points $\Phi_{control} = 0,\:-1,\:-2.5\,\mbox{mrad}$.}
\end{figure}

In the second experiment, the pendulum is rotated by a large angle with respect to the GRS to simulate a large DC acceleration.  In the current configuration, a rotation angle \begin{math}\Phi_{EQ}\approx 2\,\mbox{mrad}\end{math} requires a DC torque of roughly $13\,\mbox{pN}\cdot\mbox{m}$ to keep the test mass centered, a differential force of roughly $1.3\,\mbox{nN}$ (with electrostatic softening of the pendulum, this gives \begin{math}\Phi_{EQ}\approx 5.3\,\mbox{mrad}\end{math}, see Figure \ref{fig:utn_pendulum_data}). The measured angular displacement is again converted into torque using (\ref{eq:rot2torque}) and the contribution from the noisy electrostatic actuation produces an excess in noise power relative to the first configuration.

In the final measurement, the pendulum rotation of the second experiment is maintained but a free-fall control scheme is employed to control the position of the TM. Torque impulses are applied periodically with a duty cycle $\chi$, with average amplitude $N_{kick}\approx-\Gamma\Phi_{EQ}/\chi$. The free-fall torque noise can then be measured and compared with both the background and continuous-actuation cases. 

The pendulum dynamics in between two impulses is a free oscillation around the equilibrium point. The torsional spring is small and positive, in contrast to the small negative spring expected in orbit. Because the stiffnesses are small and the flight time relatively short, 350 s flight compared to the 830 s free-oscillation period, the motion is similar to a parabolic flight in both cases. 
 
The motion is periodically forced, by the impulses applied, to come back to a single initial position with a chosen velocity, to allow a periodic flight. To do that, a control scheme is implemented, where an observer estimates the pendulum position and velocity before each impulse with least squares fitting of the pendulum rotation data. Then a controller estimates the impulse intensity  needed to reach the ideal initial point for the next cycle, using the pendulum dynamic constants and the flight and impulse times. We can also vary $\Phi_{control}$, the average position of the mass during the free flight, to change the actuation level. 

Example preliminary data are shown in Figure \ref{fig:utn_pendulum_data}, with flight times of 90 and 250 s, using pendulum periods of 482 and 830 s, respectively, employing also different controller set points. The controller has been successfully employed with a variety of control configurations. The next step is that of debugging and understanding the torque noise in our on-ground free-fall model and beginning to test our flight data analysis algorithms on the experimental data.  

\section{Data Analysis Challenge}
\label{sec:DAchallenge}
At their core, the goals of the LPF drift mode experiments are the same as those conducted in the standard science mode: measurement of the spectrum of the acceleration of the RTM relative to the NTM and estimation of system parameters such as the gravity gradients, stiffness terms, etc. The general analysis strategy for LPF experiments begins with models of the expected acceleration that are parametrized by relevant system parameters. Fitting these models to the data provides estimates of the system parameters and the fit residuals can be used to estimate the relative test mass acceleration.

The drift mode data poses two unique challenges to this approach. The first is the presence of the kicks, which represent a high-noise configuration of the NTM and can't be used for spectral estimation. Consequently, they must be excised in some way. The second is the size of the free-flight signals relative to the noise levels of interest. In displacement, the free flights have an amplitude of $\sim 10\,\mu\mbox{m}$, compared with a displacement sensitivity of $7\,\mbox{pm}/\sqrt{\mbox{Hz}}$. In acceleration, the force bias on the NTM is equivalent to an acceleration of $10^{-10}\,\mbox{m}/\mbox{s}^2$, compared with expected noise levels of $\sim 10^{-14}\,\mbox{m}/\mbox{s}^2/\sqrt{\mbox{Hz}}$. Residuals in fitting the free-flight terms caused by small parameter or model errors can significantly impact the resulting estimate of the relative acceleration between the NTM and RTM. The effect of such residuals is exacerbated by the presence of the data gaps around the excised kicks, to which we now turn our attention. 

After the `deterministic' portion of the free-flight data has been removed, the next step is to estimate the spectrum of the residuals. The data from individual free-flights is well suited to this purpose and standard spectral estimation techniques can be applied. Unfortunately, the minimum Fourier frequency at which the spectrum can be estimated is determined by the length of the free flight segments. For the $\sim 350\,\mbox{s}$ flights for the LTP experiment, this limits the estimation to $f > 3\,\mbox{mHz}$. To move to the lower end of the LTP band ($1\,\mbox{mHz}$) and to the full LISA band ($0.1\,\mbox{mHz}$), requires combining data from successive flights. Since the portions of data during the kicks cannot be used, this amounts to the not-uncommon problem of estimating the underlying spectrum in the presence of gaps. The most common technique for estimating spectra in unevenly-sampled timeseries is the Lomb-Scargle method \cite{Lomb1976, Scargle1982, Press1989}, which is a mainstay in astronomical data analysis. However, Lomb Scargle is not ideally suited to the LPF free-flight problem because the data gaps are regular and periodic and the dynamic range of the spectra is larger than the typical simple power-law spectra encountered in astrophysics.

\begin{figure}[h!]
	\centering
        \subfloat[Ensemble-averaged noise spectral densities\label{fig:noiseComp}]{\includegraphics[width=7 cm]{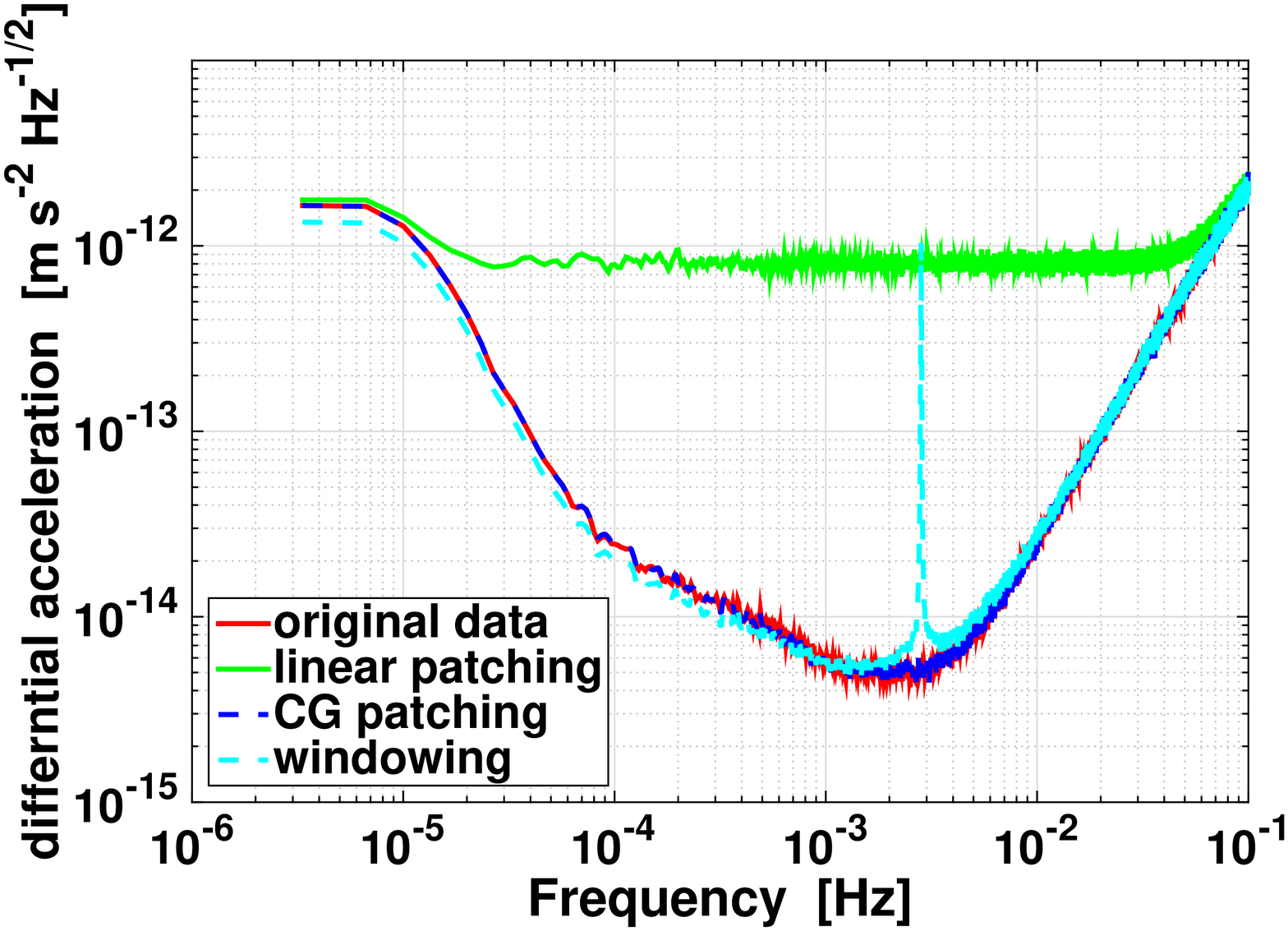}}\quad
      \subfloat[Relative Error\label{fig:errorComp}]{\includegraphics[width=7 cm]{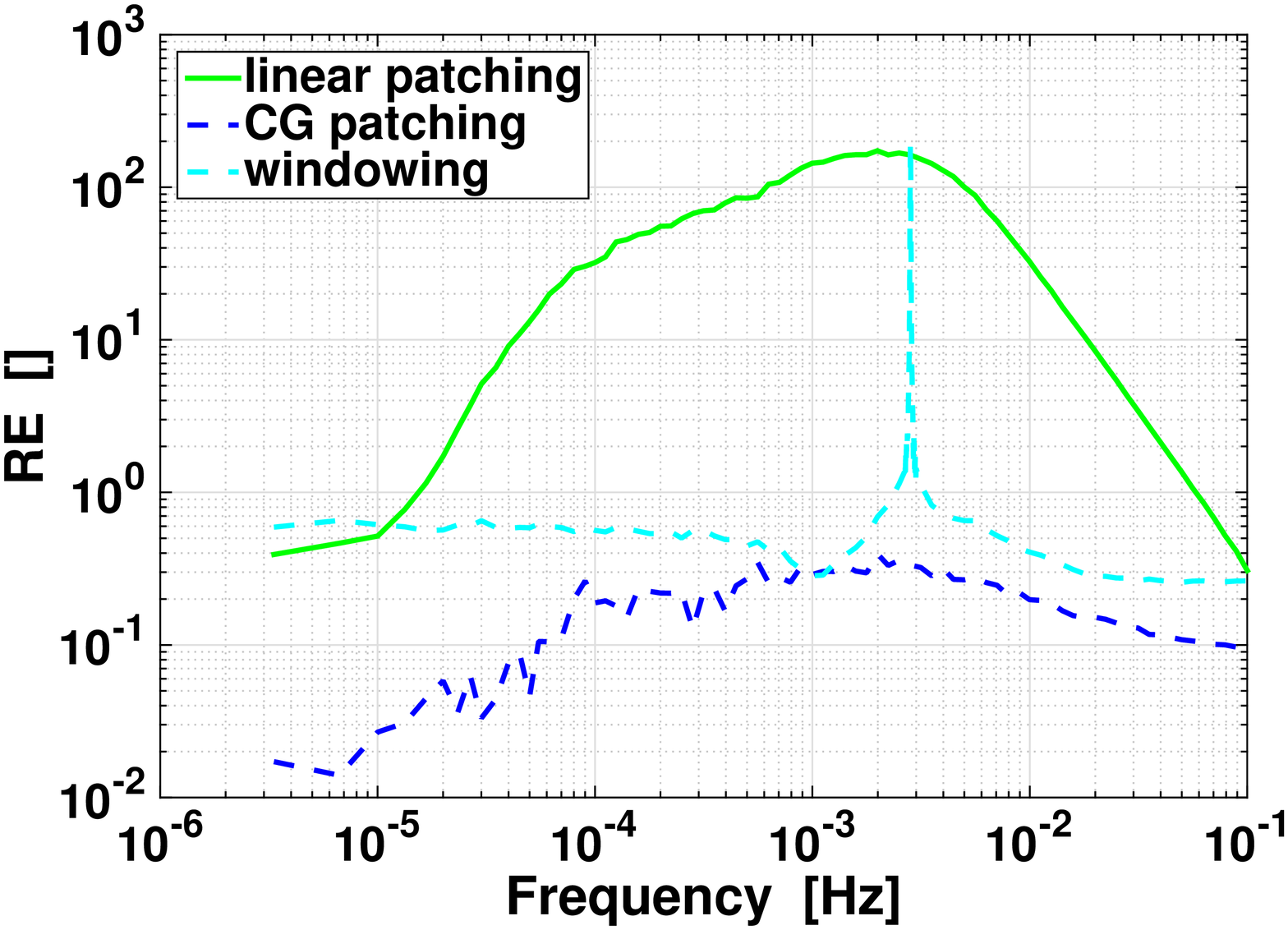}}\quad
	\caption{Comparison of methods for spectral estimation of RTM acceleration relative to NTM with data gaps of 4 s duration separated by 350s free-flight segments. A linear state-space simulator of LTP was used to generate an ensemble of 100 noise realizations of LTP science mode data with no electrostatic suspension noise and data gaps were artificially inserted allowing for the comparison of the estimated power spectral densities with and without gaps. The left plot shows the power spectral densities of the original signal (red), the linearly-patched signal (green), the Constrained-Gaussian patched signal (dashed blue), and the power spectral density estimated with the window method (dashed cyan). The right plot shows the relative error as defined in (\ref{eq:fracErr}) between the power spectral density estimated with gaps and that without gaps for the three approaches. }
\end{figure}

The general effect of data gaps is to introduce systematic biases into the estimated spectrum of the underlying continuous process.  The precise nature of this bias depends both on the characteristics of the gaps (their duration, number, and grouping) as well as on the spectrum of the signal. The primary effect on LPF free-flight data is to `fill in the bucket' in the LPF sensitivity. This is due to power from the upper end of the LPF band aliasing into lower frequencies and is exacerbated by the large range in noise power between the minimum noise of $\sim 3\times 10^{-14}\,\mbox{m}/\mbox{s}^2/\sqrt{\mbox{Hz}}$ at $f\sim 3\,\mbox{mHz}$ and the noise of $\sim 5\times 10^{-12}\,\mbox{m}/\mbox{s}^2/\sqrt{\mbox{Hz}}$ at $f\sim 100\,\mbox{mHz}$.  To illustrate this effect, we generated a series of `noise-only' mock LTP drift mode data using a  linear state-space simulator of the LTP experiment. The simulator was configured to run in nominal science mode (no free-flights) with the capacitive actuation noise of the NTM along the sensitive axis artificially turned off to mimic the noise environment in the drift mode experiment. The red trace in Figure \ref{fig:noiseComp} shows the average power spectral density of the residual RTM-NTM acceleration noise for an ensemble of 100 runs of the simulator with different noise seeds, representing the expected level in a drift mode experiment. Gaps of $4\,\mbox{s}$ duration separated by $350\,\mbox{s}$ continuous segments were then artificially placed in the data to mimic the portion of the data that would be removed to avoid the influence of the kicks during drift mode. The increase in gap size over the $1\,\mbox{s}$ kick duration allows the suppression of transients generated during downsampling of the data from $10\,\mbox{Hz}$ to $1\,\mbox{Hz}$ and an estimate of the acceleration from the measured displacement timeseries using finite differencing. As a zeroth-order estimate for the data in the gaps, a linear `patch' is placed in each gap connecting the beginning point with the end point of the gap. The spectral estimate from this linearly patched data is the green trace in Figure \ref{fig:noiseComp}. It is clear that the linearly patched data gives a severely overestimated noise level over the entire LPF band. Figure \ref{fig:errorComp} shows the fractional error of the spectral estimates over the lower portion of the LPF/LISA frequency band,
\begin{equation}
\mbox{RE}(f)\equiv\frac{\left| \overline{S_{patched}(f)-S_{original}(f)}\right|}{\sigma_{original}(f)}
\label{eq:fracErr}
\end{equation}
where $\overline{S_{original}(f) - S_{patched}(f)}$ represents the error in the power spectral density in each frequency bin for a given patching method, averaged over the ensemble of 100 simulator runs. $\sigma_{original}(f)$ is the standard deviation of the power spectral density in each frequency bin over the ensemble. $\mbox{RE}(f)$ measures the error introduced by the patching method relative to the statistical uncertainty associated with a single simulator run. If $\mbox{RE}(f) \lesssim 1$, then the error in the patching method is not significant for an individual free flight experiment. For the linear patching technique (green trace), $\mbox{RE}(f)>1$ for most of the LPF measurement band, with a maximum value of more than $10^2$ at Fourier frequencies of a few mHz.

\section{Approaches to Data Analysis}
\label{sec:DAmethods}
In this section, we briefly outline some of these techniques developed as part of this ongoing effort, with the details left to future publications.

\subsection{Windowing of data gaps}
\label{sec:windowing}

One strategy for mitigating the adverse effects of data gaps on the estimation of spectra from timeseries is to employ windowing. When estimating spectra from timeseries without gaps, it is typical to first multiply the time series by a window function that smoothly tapers to zero at the beginning and end of the timeseries. This helps to suppress artifacts caused by mismatch between the beginning and the end of the timeseries. This same strategy can be applied to data with gaps by applying a window to each individual segment of continuous data. The cyan line in Figure \ref{fig:noiseComp} shows the result of the spectral estimate made using a $sin^2$ window in each free-flight segment. The estimate of the spectra made from this windowed data is a far better approximation of the original data than that made using the linear patches. However, the windowed estimate shows a large excess relative to the original spectrum near the inverse gap separation of $1/354\,\mbox{s}\approx 2.8\,\mbox{mHz}$\footnote{To clarify, $4\,\mbox{s}$ gaps separated by $350\,\mbox{s}$ free-flight segments gives a center-to-center gap spacing of $354\,\mbox{s}$.}. With the exception of the inverse gap separation frequency, $\mbox{RE}(f)<1$ for most of the LPF measurement band.

To better compare the original acceleration noise power spectrum to the one recovered using the window method, fits were made to both versions of the power spectral density from each simulator run. The model was a four-component power law\footnote{Note that the fit was performed on the \emph{power} spectral density where as Figures \ref{fig:LPFreqs}, \ref{fig:LPFCBE}, and \ref{fig:noiseComp} plot the \emph{amplitude} spectral density. As a result the spectral indices are a factor of two greater.},
\begin{equation}
S_{mod}(f)= P_{-6}\cdot f^{-6}+P_{-2}\cdot f^{-2}+P_{0}\cdot f^{0}+P_{4}\cdot f^{+4},
\label{eq:powMod}
\end{equation}
where $f$ is the Fourier frequency and $P_{-6,-2,0,+4}$ are the amplitudes of the four components. Figures \ref{fig:histsPm6} - \ref{fig:histsPp4} show histograms of the best-fit amplitudes for the spectral model over the ensemble of 100 simulator runs for both the original data (red) and the windowed method (cyan). Table \ref{tab:fitCompare} lists the mean and standard deviations for the ensemble of noise realizations for each of the four component amplitudes. As a rough measure of the statistical equivalence of the distributions of the best-fit coefficients to the original and recovered power spectral densities, we compute the difference between the means normalized by the quadrature sum of the deviations. The window method produces significant bias in the $P_{-2}$ and $P_{0}$ coefficients (3.5 and 6 sigma, respectively) and slight bias (0.9 sigma, or a p-value of 36\%) in the $P_{+4}$ coefficient. The recovered $P_{-6}$ coefficient is consistent but is not well-determined in the fits to either the original or recovered data.

\begin{table}[h!]
\caption{\label{tab:fitCompare}Comparison of best-fit parameters for spectral model in (\ref{eq:powMod}) for original and recovered power spectral densities.}
\begin{center}
\begin{tabular}{|c|c|c|c|c|c|c|c|c|}
\hline
& \multicolumn{2}{|c|}{original} & \multicolumn{3}{|c|}{window method} & \multicolumn{3}{|c|}{CG patching} \\
& $\mu$ & $\sigma$ & $\mu$ & $\sigma$ & $\frac{\mu-\mu_{orig}}{\sqrt{\sigma^2+\sigma_{orig}^2}}$ & $\mu$ & $\sigma$ & $\frac{\mu-\mu_{orig}}{\sqrt{\sigma^2+\sigma_{orig}^2}}$ \\
\hline
$P_{-6}\times 10^{53}$ & $0.66$ & $0.66 $ &  $0.65 $ &  $0.61 $ & 0.03 &  $0.66 $ &  $0.71 $ & 0.03 \\
$P_{-2}\times 10^{35}$ & $1.14$ & $0.18 $ &  $0.44 $ &  $0.09 $ & -3.5 &  $1.08 $ &  $0.20 $ & -0.2 \\
$P_{0}\times 10^{29}$ & $2.25$ & $0.16 $ &  $4.29 $ &  $0.30 $ & 6.1 &  $2.31 $ &  $0.19 $ & 0.3 \\
$P_{+4}\times 10^{20}$ & $6.23$ & $0.03 $ &  $6.28 $ &  $0.04 $ & 0.9 &  $6.22 $ &  $0.03 $ & 0.09 \\
\hline
\end{tabular}
\end{center}
\label{default}
\end{table}%

\subsection{Constrained-Gaussian Gap Patching}
\label{sec:patching}
Another approach to dealing with data gaps and the difficulties they cause with estimating spectra is to fill the gaps with fabricated data. While the fabricated data in these patches will necessarily bias the spectral estimate, if the gaps are not too numerous and the patch data are chosen carefully, the bias can be far less than what arises from the gaps themselves. To minimize the bias resulting from the patches, the data in the patches should have the same spectral content as the missing data it is replacing. The first requirement is that the data points in the patches have proper correlations with \emph{one another}, this is relatively straightforward problem of drawing random samples of (unevenly-sampled) data with a corresponding spectrum.  The second requirement is that the data points in the patches have proper correlation with the \emph{existing data}. In Appendix A, we demonstrate that enforcing the correlation with the existing data can be accomplished by biasing the mean values of the random data used to make the patches, a technique we call \emph{Constrained-Gaussian Gap Patching} or CG patching. The values for these biases can be entirely computed from the existing data and a model of the underlying spectrum.

\begin{figure}[h!]
	\centering
	\subfloat[$f^{-6}$ component\label{fig:histsPm6}]{\includegraphics[width=7 cm]{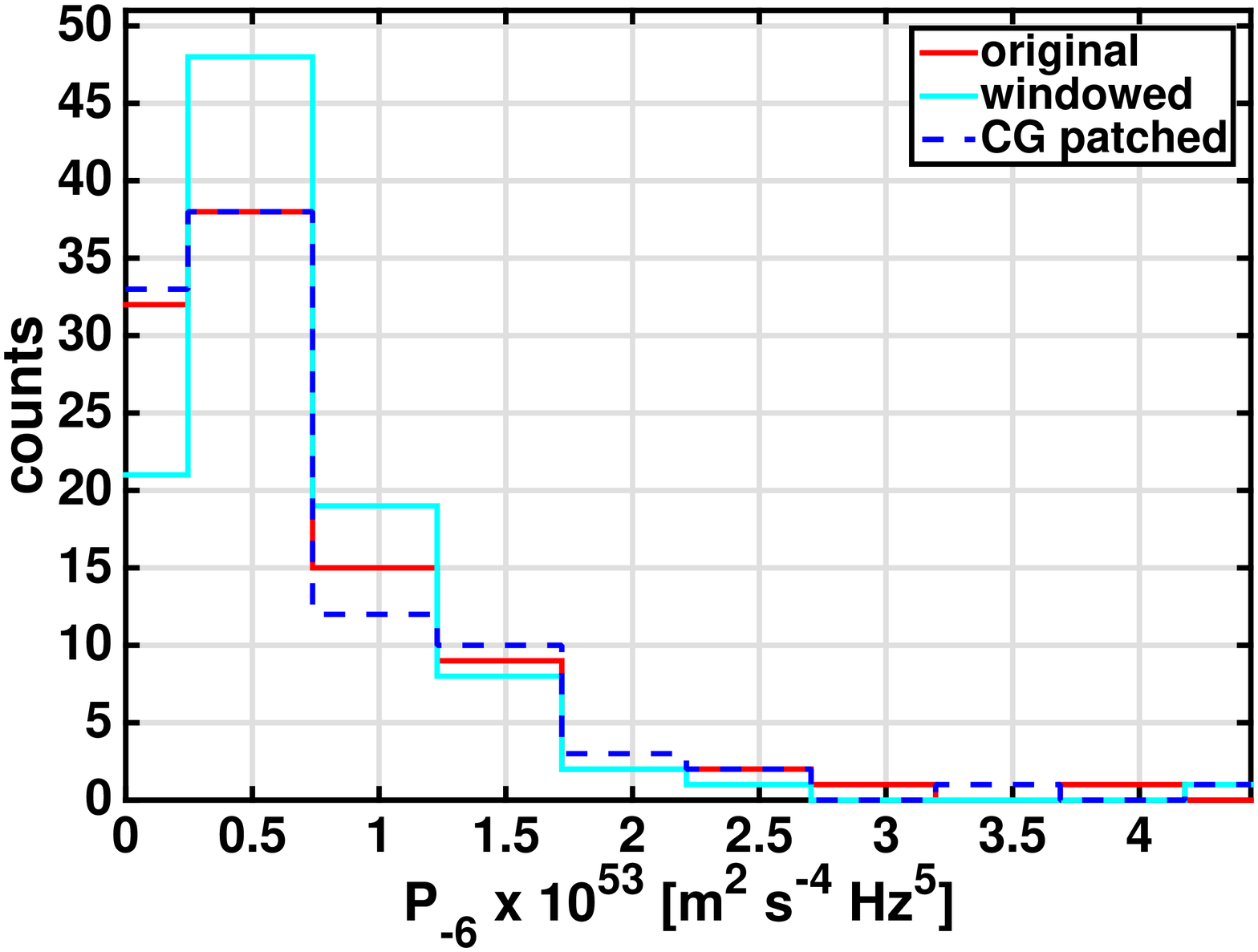}}\quad
	\subfloat[$f^{-2}$ component\label{fig:histsPm2}]{\includegraphics[width=7 cm]{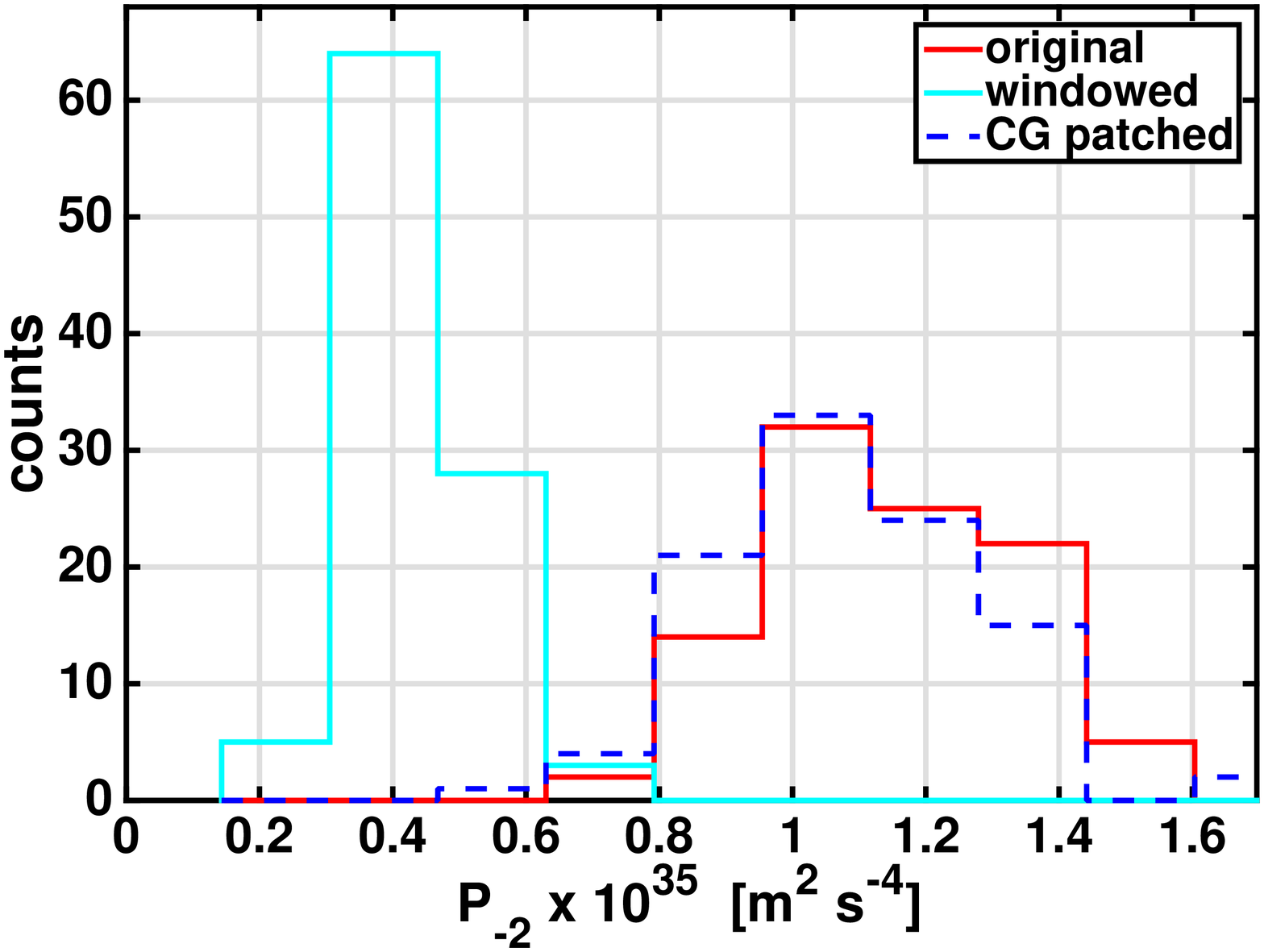}}\\	
         \subfloat[$f^{0}$ component\label{fig:histsP0}]{\includegraphics[width=7 cm]{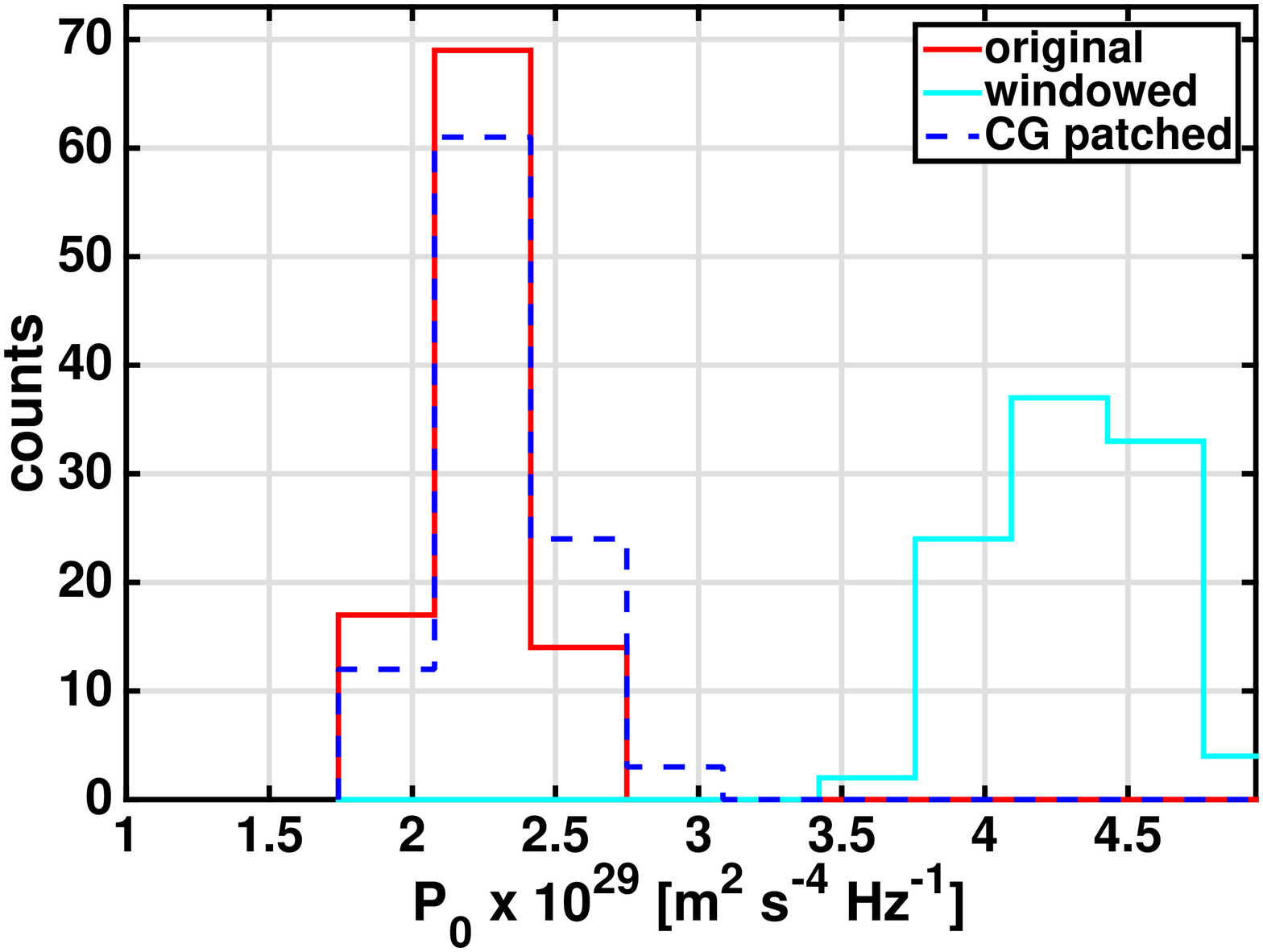}}\quad
	\subfloat[$f^{+4}$ component\label{fig:histsPp4}]{\includegraphics[width=7 cm]{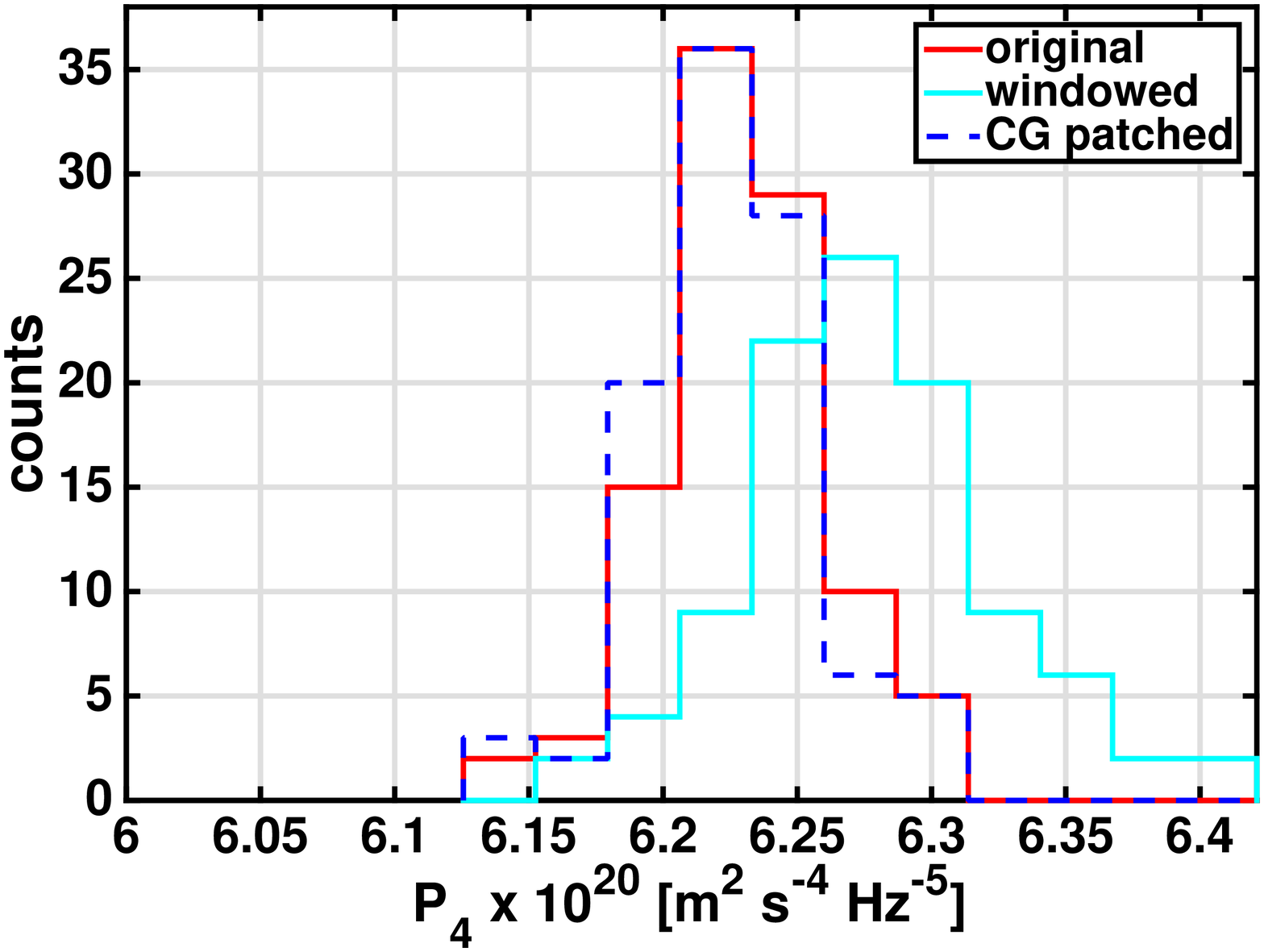}}\\	
	\caption{Histograms of power-law amplitudes in a four-component fit to the spectrum of RTM acceleration relative to the NTM for the original data (red) and as recovered in the presence of gaps using the window method (cyan) and the Constrained  Gaussian patching method (blue dashed). For each of the 100 simulator runs, a power spectrum was generated and fit using a least-squares algorithm to a four-component power law with indices -6,-2,0, and +4 as described in (\ref{eq:powMod}). The plots show the distribution of best-fit parameters over the ensemble of 100 simulator runs. For the case of the windowed method, the noise spike around $2.8\,\mbox{mHz}$ was de-weighted to minimize it's bias on the fit.}
\end{figure}

Once patched, standard approaches can be used to compute the spectrum of the entire data set as if it were continuous, without suffering from the bias introduced by the gaps. In principal this is a catch-22, one must know the spectrum in order to pick the patches that will allow the spectrum to be measured. In practice, an approximate \emph{a priori} model of the spectrum will suffice. If necessary, an iterative procedure can be used in which an initial guess for $S_y$ is used to generate the patches, a new value for $S_y$ is estimated from the patched data series, and that spectrum is used to determine new patches. 

The CG-patching approach was applied to the mock `noise-only' drift-mode data in Figure \ref{fig:noiseComp}. The ensemble-average spectra is shown in the blue dashed trace in Figure \ref{fig:noiseComp} and matches the average spectra from the original data (red) quite closely for all frequencies.  The value of $\mbox{RE}(f)$ is less than unity across the entire measurement band and is significantly less than that of the windowed method for most frequencies. More significantly, when the spectra recovered with CG patching are fit to the power-law model in (\ref{eq:powMod}), the distribution in amplitudes is quite close to that of the original data, as shown in the blue dashed traces in the histograms in Figures \ref{fig:histsPm6} - \ref{fig:histsPp4}.  Table \ref{tab:fitCompare} includes a statistical comparison between the fits to the power spectral densities of the original and CG-patched data. The normalized difference between the distributions in fit amplitudes is less than 0.3 sigma for all coefficients.

\subsection{Other methods}
\label{sec:others}
While the window and CG-patching methods have received the most attention for potential application to data from free-flight experiments, other methods are also being studied. One approach is to recognize that if an approximate model for the underlying spectra is known (as is also required for the CG-patching method) and the location of the gaps are known, the spectral bias induced by the gaps can be analytically removed. This is a classic example of a deconvolution problem with the associated computational difficulties. However, if the data is aggressively low-pass filtered, the problem becomes tractable. 

An altogether unique approach is to consider only the data in the free-flights themselves and perform time-domain fits on the free-flight trajectories. Each of these fits will have parameters such as residual gravity gradient that are the same as those for the first stage, explained at the beginning of Section \ref{sec:DAchallenge}, of the methods outlined above. However, there will also be some variation in these parameters between successive flights that is caused by low-frequency (Fourier frequencies lower than the inverse free-flight time) acceleration noise in the system. By analyzing the variation in parameters recovered from fits to a series of free-flight trajectories, an estimate of acceleration noise at Fourier frequencies below the inverse free-flight frequency can be made without the need to deal with data gaps.

\section{Conclusions}
\label{sec:conclusions}

The free fight experiments planned for LPF will provide an opportunity to gather data that is even more representative of a full-scale LISA like instrument than the standard LPF science mode. This improvement will allow for better characterization of the small forces that ultimately limit the performance of LPF, LISA, and other future instruments requiring low-disturbance environments such as advanced geodesy or fundamental physics missions. The challenges in extracting this information from the free-flight data are significant, but progress in overcoming them is being made with a number of independent techniques. The challenge of estimating the relative RTM/NTM acceleration noise in the presence	of data gaps has been successfully addressed by a number of techniques with the current limitation being the ability to combine this with accurate removal of the deterministic free-flight signal. It is fully expected that the remaining challenges will be resolved in time for launch and operations.

\section*{Appendix A: Generating Constrained-Gaussian Patches}
\label{sec:CGfill}
In this appendix, we introduce the mathematical formalism of constrained-Gaussian gap patching. This method is similar to techniques used to patch gaps in 2D sky maps of cosmic microwave background data \cite{Hoffman1991, Bucher2011}. In the time domain, spectral information is encoded in the two-point correlation function, $C^{jk}$, which measures the correlation between the samples $y^j$ and $y^k$. For stationary processes, the two-point function depends only on the separation between the samples,  $C^{jk}=\mathcal{C}(j-k)$, and is directly related to the spectral density,

\begin{equation}
C^{jk} = \int_{-\infty}^{+\infty}{\frac{1}{2}S_y(f)e^{-2\pi fT(j-k)}}.
\label{eq:Cjk_def}
\end{equation}  

Under the further assumption that $y^j$ measures a Gaussian random process, the two-point function can be used to express the probability distribution function (PDF) for $y^j$,

\begin{equation}
P(\vec{y})\propto\exp\left[-\frac{1}{2}\sum_{j,k}{C_{jk}y^j y^k}\right],
\label{eq:probDist}
\end{equation}
where $C_{jk}$ is the inverse of $C^{jk}$ such that $C_{jl}C^{lk}=\delta^{j}_{k}$.

For the case of data with gaps, it is useful to separate the data $y_j$ into portions inside the gaps, $j\in\G$, and data between the gaps $j\in\Gbar$. The sum in (\ref{eq:probDist}) can then be divided into three parts,

\begin{equation}
\ln P(\vec{y}) = const - \frac{1}{2}\left[\sum^{\Gbar,\Gbar '}{C_{jk} y^j y^k} + 2\sum^{\Gbar,\G}{C_{jk} y^j y^k} + \sum^{\G,\G '}{C_{jk} y^j y^k}\right].
\label{eq:logProbDist}
\end{equation}

The first term in (\ref{eq:logProbDist}) is a fixed constant depending on measured data that can be absorbed into the overall normalization constant. The final term depends entirely on the data within the gaps, and just describes the PDF for (unevenly sampled) Gaussian noise with a particular spectrum.  The middle term encodes the relationship between data inside the gaps and data outside the gaps. Faithfully reproducing that relationship is the key to generating good patches for filling gaps in measured data.
Since we are interested in generating the data within the gaps, we can factor the summation of the middle term in (\ref{eq:logProbDist}) and define $\lambda_j \equiv \sum_j^{\Gbar}{C_{jk}y^j}$ such that that term becomes $2\sum^\G{\lambda_j y^j}$. The PDF can then be re-written as

\begin{equation}
\ln P(\vec{y})=const-\frac{1}{2}\sum^{\G,\G '}{C_{jk}(y^j-\Delta^j)(y^k-\Delta^k)},
\label{eq:probDistSquare}
\end{equation}
where $\Delta^j\equiv -\sum_k^{\Gbar}{C^{jk}\lambda_k}$. Comparing (\ref{eq:probDistSquare}) with the standard expression for multivariate Gaussian distributions, one finds that the data in the gap is described by a multivariate Gaussian with covariance $C_{jk}$ and non-zero means $\Delta^k$. Note that the values for $\Delta^k$ are entirely determined by the data outside the gap. 

The prescription for \emph{Constrained Gaussian Gap Filling} is thus as follows: make a guess for spectral density $S_y(f)$ describing the data; compute an estimated two-point correlation function, $C_{jk}$; compute the $\Delta^k$ for each point on the gap based on the data outside the gap and the estimated $C_{jk}$, draw data from a multivariate Gaussian distribution with covariance $C_{jk}$ and mean $\Delta^k$.

\section*{References}
\bibliographystyle{iopart-num}
\bibliography{bibliography}

\providecommand{\newblock}{}
\begin{thebibliography}{10}
\expandafter\ifx\csname url\endcsname\relax
  \def\url#1{{\tt #1}}\fi
\expandafter\ifx\csname urlprefix\endcsname\relax\def\urlprefix{URL }\fi
\providecommand{\eprint}[2][]{\url{#2}}

\bibitem{Bondi1959}
{Bondi} H, {Pirani} F~A~E and {Robinson} I 1959 {\em Royal Society of London
  Proceedings Series A\/} {\bf 251} 519--533

\bibitem{Armano_09}
Armano M and et~al 2009 {\em Class. Quant. Grav.\/} {\bf 26}

\bibitem{Lange1964}
Lange B 1964 {\em AIAA Journal\/} {\bf 2} 1590--1606
  \urlprefix\url{http://dx.doi.org/10.2514/3.55086}

\bibitem{DeBra1997}
deBra D~B 1997 {\em Classical and Quantum Gravity\/} {\bf 14} 1549
  \urlprefix\url{http://stacks.iop.org/0264-9381/14/i=6/a=026}

\bibitem{Antonucci2011}
Antonucci F, Armano M, Audley H, Auger G, Benedetti M, Binetruy P, Boatella C,
  Bogenstahl J, Bortoluzzi D, Bosetti P, Brandt N, Caleno M, Cavalleri A, Cesa
  M, Chmeissani M, Ciani G, Conchillo A, Congedo G, Cristofolini I, Cruise M,
  Danzmann K, Marchi F~D, Diaz-Aguilo M, Diepholz I, Dixon G, Dolesi R, Dunbar
  N, Fauste J, Ferraioli L, Fertin D, Fichter W, Fitzsimons E, Freschi M, Marin
  A~G, Marirrodriga C~G, Gerndt R, Gesa L, Giardini D, Gibert F, Grimani C,
  Grynagier A, Guillaume B, Cervantes F~G, Harrison I, Heinzel G, Hewitson M,
  Hollington D, Hough J, Hoyland D, Hueller M, Huesler J, Jeannin O, Jennrich
  O, Jetzer P, Johlander B, Killow C, Llamas X, Lloro I, Lobo A,
  Maarschalkerweerd R, Madden S, Mance D, Mateos I, McNamara P~W, Mendes J,
  Mitchell E, Monsky A, Nicolini D, Nicolodi D, Nofrarias M, Pedersen F,
  Perreur-Lloyd M, Perreca A, Plagnol E, Prat P, Racca G~D, Rais B,
  Ramos-Castro J, Reiche J, Perez J~A~R, Robertson D, Rozemeijer H, Sanjuan J,
  Schleicher A, Schulte M, Shaul D, Stagnaro L, Strandmoe S, Steier F, Sumner
  T~J, Taylor A, Texier D, Trenkel C, Tombolato D, Vitale S, Wanner G, Ward H,
  Waschke S, Wass P, Weber W~J and Zweifel P 2011 {\em Classical and Quantum
  Gravity\/} {\bf 28} 094002
  \urlprefix\url{http://stacks.iop.org/0264-9381/28/i=9/a=094002}

\bibitem{Audley2011}
Audley H, Danzmann K, Marin A~G, Heinzel G, Monsky A, Nofrarias M, Steier F,
  Gerardi D, Gerndt R, Hechenblaikner G, Johann U, Luetzow-Wentzky P, Wand V,
  Antonucci F, Armano M, Auger G, Benedetti M, Binetruy P, Boatella C,
  Bogenstahl J, Bortoluzzi D, Bosetti P, Caleno M, Cavalleri A, Cesa M,
  Chmeissani M, Ciani G, Conchillo A, Congedo G, Cristofolini I, Cruise M,
  Marchi F~D, Diaz-Aguilo M, Diepholz I, Dixon G, Dolesi R, Fauste J, Ferraioli
  L, Fertin D, Fichter W, Fitzsimons E, Freschi M, Marirrodriga C~G, Gesa L,
  Gibert F, Giardini D, Grimani C, Grynagier A, Guillaume B, Cervantes F~G,
  Harrison I, Hewitson M, Hollington D, Hough J, Hoyland D, Hueller M, Huesler
  J, Jeannin O, Jennrich O, Jetzer P, Johlander B, Killow C, Llamas X, Lloro I,
  Lobo A, Maarschalkerweerd R, Madden S, Mance D, Mateos I, McNamara P~W,
  Mendes J, Mitchell E, Nicolini D, Nicolodi D, Pedersen F, Perreur-Lloyd M,
  Perreca A, Plagnol E, Prat P, Racca G~D, Rais B, Ramos-Castro J, Reiche J,
  Perez J~A~R, Robertson D, Rozemeijer H, Sanjuan J, Schulte M, Shaul D,
  Stagnaro L, Strandmoe S, Sumner T~J, Taylor A, Texier D, Trenkel C, Tombolato
  D, Vitale S, Wanner G, Ward H, Waschke S, Wass P, Weber W~J and Zweifel P
  2011 {\em Classical and Quantum Gravity\/} {\bf 28} 094003
  \urlprefix\url{http://stacks.iop.org/0264-9381/28/i=9/a=094003}

\bibitem{Carbone2007}
Carbone L, Ciani G, Dolesi R, Hueller M, Tombolato D, Vitale S, Weber W~J and
  Cavalleri A 2007 {\em Phys. Rev. D\/} {\bf 75}(4) 042001
  \urlprefix\url{http://link.aps.org/doi/10.1103/PhysRevD.75.042001}

\bibitem{Cavalleri2009}
Cavalleri A, Ciani G, Dolesi R, Hueller M, Nicolodi D, Tombolato D, Wass P~J,
  Weber W~J, Vitale S and Carbone L 2009 {\em Classical and Quantum Gravity\/}
  {\bf 26} 094012
  \urlprefix\url{http://stacks.iop.org/0264-9381/26/i=9/a=094012}

\bibitem{Grynagier10}
Grynagier A 2010 The drift mode for {LISA} {P}athfinder, control description
  Tech. Rep. S2-iFR-TN-3005 Universit\"at Stuttgart

\bibitem{Lomb1976}
{Lomb} N~R 1976 {\em Astrophysics and Space Science\/} {\bf 39} 447--462

\bibitem{Scargle1982}
{Scargle} J~D 1982 {\em The Astrophysical Journal\/} {\bf 263} 835--853

\bibitem{Press1989}
{Press} W~H and {Rybicki} G~B 1989 {\em The Astrophysical Journal\/} {\bf 338}
  277--280

\bibitem{Hoffman1991}
Hoffman Y and Ribak E 1991 {\em The Astrophysical Journal\/} {\bf 380} L5--L8

\bibitem{Bucher2011}
{Bucher} M and {Louis} T 2012 {\em Monthly Notices of the Royal Astronomical
  Society\/} {\bf 424} 1694--1713

\end{thebibliography}

\end{document}